# Phonon thermal transport in $\beta$-NX (X=P, As, Sb) monolayers: a first-principles study of the interplay between harmonic and anharmonic phonon properties

Armin Taheri, Carlos Da Silva, Cristina H. Amon

(armin.taheri@mail.utoronto.ca)

*Department of Mechanical and Industrial Engineering, University of Toronto, Ontario, Canada, M5S 3G8*

## Abstract

The investigation of thermal properties of recently emerged two-dimensional (2D) materials is a necessary step towards fulfilling their potential applications in nano-electronics devices. In this study, the thermal conductivity of novel $\beta$-NX (X=P, As, Sb) monolayers are investigated using a first-principles density functional theory (DFT) study based on the full solution of the linearized Peierls-Boltzmann transport equation (PBTE). The results show that the room temperature thermal conductivities of $\beta$-NP, $\beta$-NAs, and $\beta$-NSb are about 1.1, 5.5, and 34.0 times higher than those of single-element $\beta$-P, $\beta$-As, and $\beta$-Sb monolayers, respectively. The phonon transport analysis reveals that higher phonon group velocities as well as phonon lifetimes are responsible for such an enhancement in the lattice thermal conductivities of $\beta$-NX (X=P, As, Sb) binary compounds compared to single-element group-VA monolayers. We found that $\beta$-NP has the minimum thermal conductivity among $\beta$-NX (X=P, As, Sb) monolayers, while it has the minimum average atomic mass, which is in contrast with the common assumption that lower mass systems exhibit higher thermal conductivities. This work demonstrates the trade-off between harmonic and anharmonic phonon properties in determining the variation of the thermal conductivity among $\beta$-NX (X=P, As, Sb) monolayers. The higher anharmonicity in $\beta$-NP is found to be responsible for the lower thermal conductivity of this monolayer.

# 1-Introduction

Triggered by the appealing electrical, optical, mechanical, and thermal properties of phosphorene (P) [1,2], two-dimensional (2D) nanostructures of group-VA such as P, As, Bi and Sb have recently been brought into focus in the research community. As opposed to gapless graphene, whose application is challenging in the semiconductor industry, 2D materials of group-VA have a direct electronic band gap, of up to about 2.62 eV [3]. Also, members of this family possess a superior electron carrier mobility that can be as high as $1000\ cm^2V^{-1}s^{-1}$ [3,4], which makes them attractive candidates for applications in next-generation circuits, electronic and optoelectronic devices. Having a variety of stable honeycomb allotropes, commonly named as $\alpha$-, β-, γ-, and δ-phase, is another distinguished feature of these monolayers [4–6]. The β-phase (buckled) and $\alpha$-phase (puckered) are generally the two most stable allotropes, with the minimum average binding energy [4]. Experimental studies have revealed that both of these phases can be synthesized using different methods such as mechanical exfoliation [7,8] or chemical vapor deposition (CVD) [9]. For instance, GaN(001) [10] can be used as a substrate to synthesize stable layers of buckled phosphorene. From a thermal perspective, the conductivity of materials in this group monotonically decreases from about 110 W/mK to 4.0 W/mK as the atomic mass increases from P to Bi [11–14]. Thus, they can be categorized as relatively low thermal conductive nanosheets compared to graphene, making them promising for thermoelectric applications. Based on first-principles density functional theory (DFT) , density functional perturbation theory (DFPT) calculations, and the full solution of the linearized Peierls-Boltzmann transport equation (PBTE), the room temperature (RM) thermal conductivity of β-P is reported to be 78 W/mK by Jain and McGaughey [11], and 108.8 W/mK by Zheng et al. [12]. Discrepancies can generally be attributed to different DFT settings and the choice for the thickness of the monolayer adopted in the study. Highly anisotropic and orthogonal thermal conductivity is a well-known property of $\alpha$-phase of group-VA monolayers, which stems from their anisotropic layered structure. The room temperature thermal conductivity of $\alpha$-P ($\alpha$-As) in zigzag direction is reported to be 110 (30.4) W/mK, almost three (four) times higher than its corresponding value in the armchair direction, 36 (7.8) W/mK [11,14]. Using the same methodology, values of about 65.4 W/mK [12], 13.8 W/mK [12], and 4.0 W/mK [13] are reported for thermal conductivity of β-As, β-Sb, and β-Bi at RT, respectively.

In addition to single-element monolayers of group-VA, the recent emergence of their stable binary compounds with tuned electrical and thermal properties has opened up a broad avenue of nanoelectronic and thermoelectric applications [15,16]. A recent theoretical study by Xiao et al. [15] revealed that there are 26 stable 2D binary alloys in which their components are from elements of group-VA, including P, As, Bi, and Sb. Based on their results, all of these nanostructures are semiconductors with electronic band gap ranging from 0.06 eV to 2.52 eV. Further studies show that group-VA binary compounds can have tailored properties compared to the single-element monolayers [17–20]. For example, charge mobility of the $\alpha$-AsP monolayer is about three times higher than that of $\alpha$-P [17,18], but its thermal conductivity is lower than single-element $\alpha$-P [20], very intriguing for thermoelectric devices. Also, the thermal conductivity of the β-SbAs monolayer is reported to be lower than both Sb and As monolayers [19]. The underlying reason is attributed to a lower phonon lifetime of β-SbAs compared to Sb and As.

Very recently, *ab initio* theoretical studies have shown that 2D binary β-NX (X=P, As, Sb, Bi) compounds with N being nitrogen are also stable in different phases such as β, $\alpha$ and δ [21,22]. Nitrogen, the lightest member of group-VA, is considered a wide-range bandgap semiconductor which also has a stable buckled 2D allotrope [23]. Ma et al. [22] proposed a possible growth of β- and $\alpha$-PN on Ag(111) and Ag(110) substrates using the CVD method which can accelerate the prospective experimental studies on β-NX (X=P, As, Sb, Bi) 2D nanosheets. It is also reported that β-NX (X=P, As, Sb, Bi) monolayers can display direct electronic bandgap in the δ-phase, while they exhibit indirect bandgap in the β- and $\alpha$-phases [21]. This bandgap can be highly tuned by mechanical in-plane strain [22], or point defects [24]. Regarding their mechanical properties, in some phases, they show a negative Poisson's ratio (NPR) [21], which results in features such as enhanced sound and vibration absorption, along with high indentation resistance and fracture toughness [25]. Apart from these unique properties, it is known that the P-N junction is a key building block in modern electronics [16]. These features allow novel β-NX (X=P, As, Sb, Bi) 2D compounds to have additional applications compared to conventional 2D materials in different industries such as aerospace [26] and medicine [27]. However, to fulfill these potential applications, having a comprehensive description of the thermal properties and phonon transport in NX (X=P, As, Sb, Bi) 2D materials is of paramount importance, which to the best of our knowledge is missing in the literature. Thus, in this study we conduct a systematic first-principles study on the thermal properties of β-NX binary compounds of the first three members of this family with a buckled structure; including β-NP, β-NAs and β-NSb. The β-phase is the preferred choice because its small lattice-misfit with the Ag substrate [22] can result in a higher experimental realization chance. We are particularly interested in elucidating: (i) how do the thermal conductivities of β-NX (X=P, As, Sb) compounds compare with those of their single-element counterparts β-X? and (ii) what is the interplay between harmonic phonon properties, average atomic masses, and anharmonicities of the lattice? This study aims to further our understanding of phonon transport in 2D materials, specifically 2D binary compounds. From this point forwards, we will use the nomenclatures β-NX and β-X to refer to the group of binary compound monolayers and single-element monolayers, respectively, with X being P, As or Sb. After this introduction, the rest of this paper is organized as follows: Section 2 discusses the computational approach and details of the DFT simulations, Section 3 presents our main results and discussions. Finally, Section 4 summarizes the main findings.

## 2- Computational Approach

We solve the PBTE using an iterative self-consistent approach as implemented in the *ShengBTE* package [28], following the thermal conductivity prediction framework discussed in [29] and [30]. Within this framework, the only required inputs are the sets of harmonic (second-order) and anharmonic (third-order) interatomic force constants (IFCs), which are obtained from DFT and DFPT calculations. Having the sets of IFCs, the thermal conductivity along the α direction can be written as

$$k_{\alpha\alpha} = \frac{1}{k_B T^2 A t N} \sum_{\lambda(k,p)} n_\lambda^0 (n_\lambda^0 + 1)(\hbar \omega_\lambda)^2 v_\lambda^\alpha v_\lambda^\alpha \tau_\lambda \ , \tag{1}$$

where $\lambda(\mathbf{k}, p)$ denotes a phonon mode with $\mathbf{k}$ being the wave vector and $p$ the phonon branch number, $A$ is the unit cell area, $t$ is the thickness of the unit cell, $\omega$ is the phonon frequency, $N$ is the number of **q** points in the first Brillouin zone (BZ) for a $\Gamma$- centered regular grid, T represents the temperature, $v_\lambda^\alpha$ is the phonon group velocity in the direction $\alpha$, $n_\lambda^0$ is the phonon occupation number given by the Bose-Einstein distribution. Also, $\tau_\lambda$ is the phonon lifetime considering the three-phonon scattering process, given by

$$\frac{1}{\tau_\lambda} = \frac{1}{N} \left( \sum_{\lambda'\lambda''}^+ \Gamma_{\lambda\lambda'\lambda''}^+ + \frac{1}{2} \sum_{\lambda'\lambda''}^- \Gamma_{\lambda\lambda'\lambda''}^- \right), \tag{2}$$

where $\Gamma^{+}_{\lambda\lambda'\lambda''}$ and $\Gamma^{-}_{\lambda\lambda'\lambda''}$ are the scattering rates for the absorption and emission three-phonon scattering processes, respectively.

In this work, we consider idealized binary compound monolayers consisting of N and X (P, As or Sb), where the atoms are arranged on a perfect non-planar honeycomb crystal without any point defect. First, the 2-atom primitive unit cells of all 2D nanostructures studied here were optimized using the QUANTUM ESPRESSO package [31] to obtain the relaxed lattice parameters. The lattice structure parameters of the β-phase are shown in Figure 1. A Perdew-Burke-Ernzerhof (PBE) exchange-correlation functional with a projected augmented wave (PAW) pseudopotential [32] is used for all DFT calculations, typical choices for 2D materials that are expected to provide the most accurate anharmonic properties [33]. The convergence thresholds on total energy and forces were chosen to be $10^{-11}$ a.u. and $10^{-10}$ a.u., respectively. We used an electronic wave-vector grid of $31 \times 31 \times 1$ for the Brillouin zone (*BZ*) sampling. The plane-wave energy cutoff is set to a value of 80Ry. The relaxed lattice parameters obtained from structural optimization are summarized in Table 1. Our results are in good agreement with those previously reported by [21] and [22], as the maximum difference is less than 0.8%. A fine phonon wave-vector grid of $35 \times 35 \times 1$ is used for the phonon calculation. For the anharmonic calculations, we used the *thirdorder.py* script [28] to generate a set of $5 \times 5 \times 1$ displaced supercell configurations, with interactions considered up to the tenth nearest neighbour. A Γ-centered Monkhorst-Pack [34] electronic wave-vector grid of $5 \times 5 \times 1$ is also used for integration over the *BZ* of the supercells. Finally, a **q** grid mesh of $80 \times 80 \times 1$ is adopted for thermal conductivity calculations and related phonon properties.

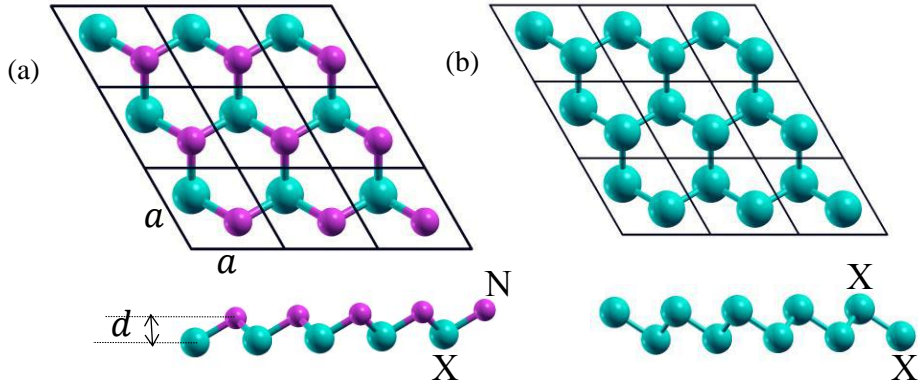

**Fig. 1.** Top and side view of monolayers (a) $\beta$-NX, and (b) $\beta$-X.

Based on Eq. (1), the thermal conductivity of a thin layer system depends on its thickness, $t$. However, this value is ill-defined in these systems [35,36]. To have a fair comparison between different monolayers, we consider the length of the unit cell along the *z*-direction ($t = 20$ Å) as the thickness of the system. Considering this length as the thickness was also suggested in [35] and [37]. For comparisons with other works, one can use the concept of thermal sheet conductance $G_s$ as $G_s = k.t$ [36].

Table 1. Average atomic mass ($\bar{M}$), lattice constant ($a$), buckling distance ($d$), scaled a-o gap, mode contribution in thermal conductivity, and thermal sheet conductance ($G_s$) for each system.

| Name | $\bar{M}(amu)$ | $a$(Å) | $d$(Å) | Scaled a-o gap | %ZA | %TA | %LA | %Optical | $G_s(WK^{-1})$ |
|---|---|---|---|---|---|---|---|---|---|
| NP | 22.49 | 2.73 | 0.86 | 0.40 | 37 | 27 | 35 | 1 | 572.3 |
| P | 30.97 | 3.28 | 1.24 | 0.62 | 57 | 23 | 17 | 3 | 518.7 |
| NAs | 44.46 | 2.98 | 0.97 | 0.94 | 29 | 24 | 46 | 1 | 769.2 |
| As | 74.92 | 3.61 | 1.39 | 0.81 | 55 | 22 | 19 | 4 | 141.4 |
| NSb | 67.88 | 3.28 | 1.02 | 1.28 | 39 | 25 | 34 | 2 | 693.2 |
| Sb | 121.76 | 4.11 | 1.64 | 1.17 | 42 | 27 | 25 | 6 | 20.4 |

# 3- Results and discussion

## 3.1- Thermal conductivity

The calculated thermal conductivities obtained from the iterative solution of the PBTE in a temperature range from 200K to 800K are plotted in Figure 2. Also, the room temperature thermal sheet conductances are listed in Table 1. All results correspond to monolayers with the same thickness of 20Å. First, we confirm that the predicted room temperature thermal sheet conductances of P (518.7 W/K), As (141.4 W/K) and Sb (20.4 W/K) are within the range of previously reported values for P (323.0 W/K [38] ,612.5 W/K [12],569.7 W/K [37] ), As (234.8 W/K [12],127.4 W/K [38], 183.0 W/K [37]) and Sb (29.5 W/K [38]), which verifies our computational workflow in this study. Figures 2(a)-2(c) compare the lattice thermal conductivity of $\beta$-NX compounds with their single-element pristine counterparts, $\beta$-X. As shown in Figure 2(a), at lower temperatures (T < 400K), monolayer NP exhibits higher thermal conductivities than $\beta$-P. However, for temperatures higher than 400K, $\beta$-NP and $\beta$-P exhibit approximately the same thermal conductivities. Guo and Dong [39] studied the lattice thermal conductivity of $\beta$-AsP alloy monolayer within the single-mode relaxation time approximation (RTA), and they also found that the thermal conductivity of $\beta$-AsP is similar to that of its single-element counterpart, $\beta$-As, over the whole temperature range from 200K to 1000K. The similarity of the thermal conductivities of $\beta$-AsP and $\beta$-As was attributed to the interplay between lower phonon lifetimes and higher phonon group velocities of $\beta$-AsP relative to $\beta$-As.

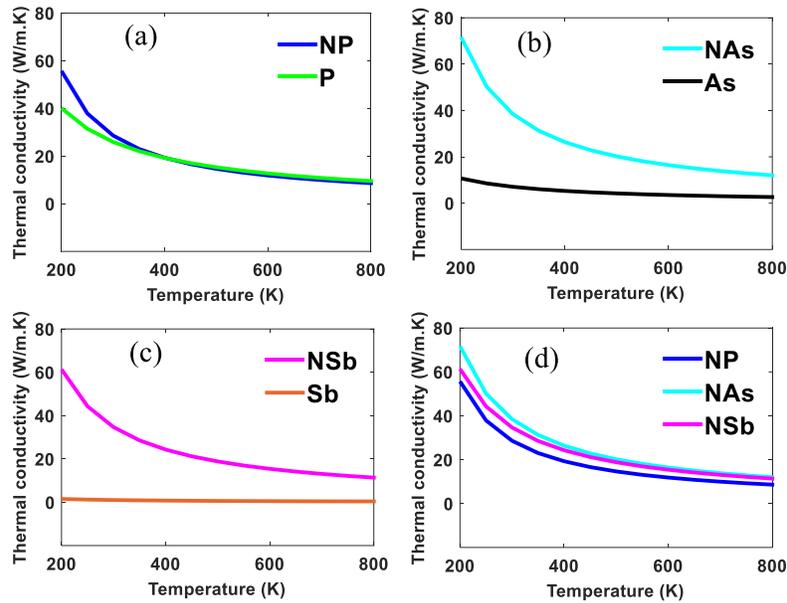

**Fig. 2.** Variation of the thermal conductivity with temperature in 20Å-thick layers of (a) $\beta$-NP and $\beta$-P, (b) $\beta$-NAs and $\beta$-As, (c) $\beta$-NSb and $\beta$-Sb, and (d) $\beta$-NP, $\beta$-NAs and $\beta$-NSb.

Based on Figure 2(b), monolayer $\beta$-NAs has higher thermal conductivity than $\beta$-As over the entire temperature range. For example, at room temperature, the thermal conductivity of $\beta$-NAs (38.46 W/m.K) is about 5.5 times higher than that of monolayer $\beta$-As (7.07 W/m.K). Similarly, Figure 2(c) shows that $\beta$-NSb exhibits a much higher thermal conductivity compared to $\beta$-Sb over the entire temperature range. We found that the room temperature thermal conductivity of a 20Å-thick layer of $\beta$-NSb is 34.66 W/m.K, almost 34.0 times higher than that of $\beta$-Sb, 1.02 W/m.K. Thus, due to their higher thermal conductivities, monolayers $\beta$-NAs and $\beta$-NSb, compared to their single-element counterparts $\beta$-As and $\beta$-Sb, are better candidates for applications requiring high heat dissipation. Finally, to better compare the thermal performance of the three binary compounds we plotted the thermal conductivity of $\beta$-NP, $\beta$-NAs and $\beta$-NSb in Figure 2(d). Our results show that $\beta$-NAs and $\beta$-NP exhibit the highest and lowest thermal conductivities over the considered temperature range, respectively. Interestingly, this is in contrast with the common assumption that materials with lower average atomic mass typically have higher thermal conductivities [40]. Similar behavior is also observed in the lattice thermal conductivity of $La_3Cu_3X_4 (X = P, As, Sb, Bi)$ and zinc blende BX (X=N, P, As, Sb) compounds reported by Pandey et al. [41] and Broido et al. [42] using the same computational approach applied in this work. Based on [41], $La_3Cu_3P_4$ has the lowest thermal conductivity compared to other compounds over a temperature range from 200K to 1000K, whereas it is the lightest mass system. Similarly, it is reported in [42] that BAs has the highest thermal conductivity between zinc blende BX (X=N, P, As, Sb) compounds for temperatures higher than 200K while its average atomic mass is higher than BN and BP. In a recent study on the variation of the thermal conductivity ($k_l$) with the average unit cell atomic mass ($\bar{M}$) in group-VA compound monolayers with the puckered structure [43], a relation of the form $k_l =$ $c_1 + \frac{c_2}{\overline{M}^2}$ is suggested for the thermal conductivity in both Zigzag and Armchair directions. Based on our results shown in Figure 3, we did not find a similar trend between $k_l$ and $\overline{M}$ in buckled monolayers of group-VA. However, to better judge the variation of the thermal conductivity with the average atomic mass in buckled group-VA monolayers, analyzing more data points is essential, which is beyond the scope of this study.

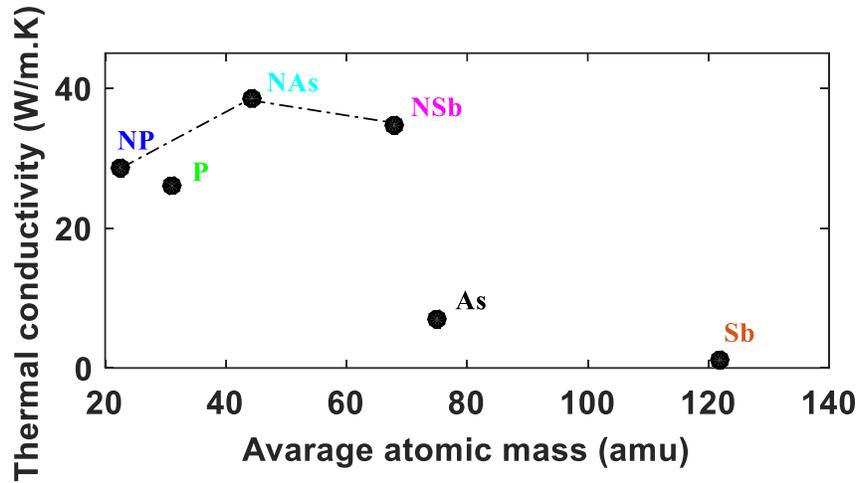

**Fig. 3.** Room temperature thermal conductivity versus the average atomic mass of unit cells for 20Å-thick layers of $\beta$-NX (X=P, As, Sb) and $\beta$-X monolayers.

Figure 4 shows the contribution of each phonon mode to the total thermal conductivity of $\beta$-NX compounds and $\beta$-X monolayers at room temperature. As shown, for all monolayers except $\beta$-NAs, the ZA mode contributes the most to the thermal conductivity. Also, we emphasize that the contribution of the ZA mode is generally higher in $\beta$-NX compounds than $\beta$-X. For instance, the contribution of the ZA, TA and LA modes in $\beta$-NP ($\beta$-P) are about 37(57) %, 27(23) % and 35(17) %, respectively. Our results are in good agreement with those of Jain and McGaughey, who reported the dominancy of the ZA mode in the thermal conductivity of $\beta$-P using the iterative solution of the PBTE [11]. However, it is worth mentioning that although the ZA mode is the dominant heat carrier in these monolayers (except $\beta$-NAs), its contribution is much lower than that of graphene, in which the ZA mode contributes to about 80% of the total thermal conductivity [44] [45]. This extremely high contribution of the ZA mode in graphene stems from the flat structure and the out-of-plane reflectional symmetry [46], which does not exist in group-VA monolayers and their alloys. Considering $\beta$-NAs, the LA mode contributes most to the phonon thermal transport, with a contribution of about 46% in the total thermal conductivity. Also, we found that in all monolayers the optical modes have negligible contribution compared to the acoustic modes. The maximum contribution of the optical modes occur in $\beta$-NSb, which still is only about 6%. This low contribution of the optical modes is usually attributed to the low group velocities and phonon lifetimes of the optical modes.

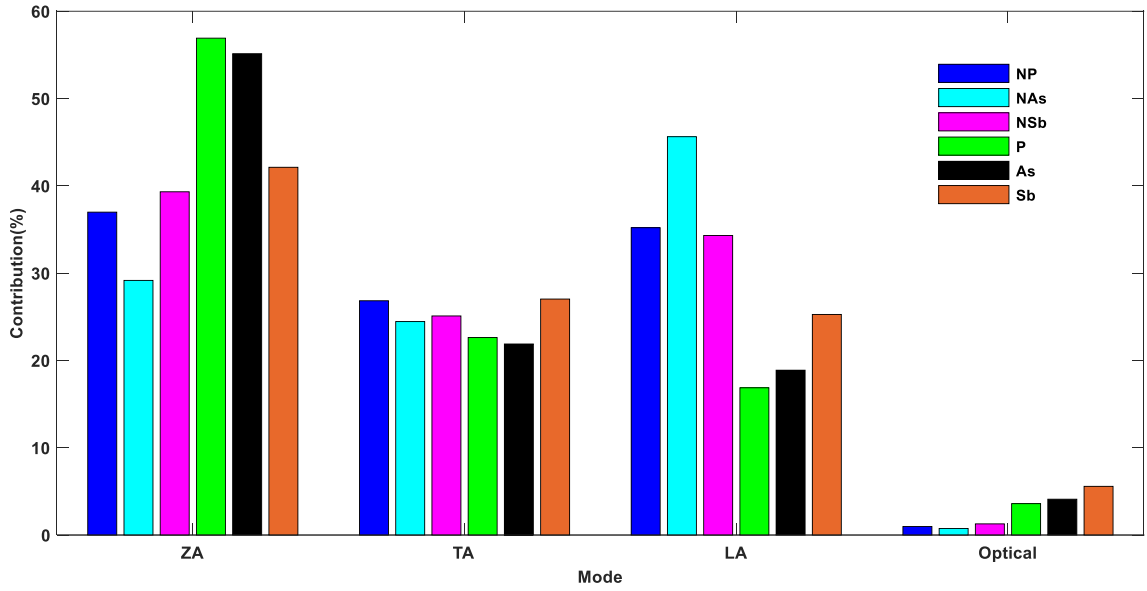

**Fig. 4.** Contribution of each phonon mode towards the total thermal conductivity in $\beta$-NX (X=P, As, Sb) and $\beta$-X monolayers at room temperature.

In order to shed light into the underlying physics of the findings discussed here, different harmonic and anharmonic phonon properties are studied in the next subsections.

## 3.2- Harmonic properties

Figure 5 shows the phonon dispersion curves of $\beta$-NX compounds along with its single-element counterpart $\beta$-X. Since all of these monolayers have two atoms per unit cell, the phonon dispersion curves consist of three acoustic and three optical branches. The out-of-plane acoustic, the in-plane transversal acoustic, and the in-plane longitudinal acoustic modes are labeled as ZA, TA and LA, respectively. It is important to note that as opposed to graphene, which has a flat lattice structure, all monolayer studies here have a buckling distance $d$, which results in the coupling of the ZA mode with the TA and LA modes. Nevertheless, we still use ZA to show the out-of-plane acoustic mode here. In the long-wavelength limit, the ZA branch of all monolayers shows a quadratic behavior which is a typical feature of 2D materials [47,48], while the two other acoustic modes are linear in the same limit.

First, we confirm that there is no imaginary frequency in the ZA branch of the dispersion curves, demonstrating that all monolayers considered here are dynamically stable. However, we emphasize that for all $\beta$-NX compound monolayers, a finer phonon wave-vector grid is required to remove all small negative frequencies near the $\Gamma$-point. In this case, a phonon wave-vector grid of $35 \times 35 \times 1$ can make the "U"-shape negative frequencies zone near the $\Gamma$-point disappear (see supplementary information). Figure 5 shows that $\beta$-NX compounds exhibit stiffer acoustic dispersion curves compared to $\beta$-X, which results in higher acoustic group velocities for $\beta$-NX. Also, the optical phonons move upwards in $\beta$-NX compounds as a result of the lower average atomic mass of the unit cell in these alloys. Phonon dispersion curves of $\beta$-AsP and $\beta$-As show similar differences as discussed in Ref. [39]. The phonon bandgap between the optical and acoustic modes (a-o gap) significantly affects the thermal conductivity by modifying the three-phonon intrinsic processes in which both acoustic and optical modes are involved, i.e., $A + A \rightarrow O$. The scaled a-o gaps, a-o bandgap width divided by the maximum acoustic frequency, are listed in Table 1, which are in excellent agreement with previously reported values [38]. Table 1 shows that $\beta$-NAs and $\beta$-NSb have higher scaled bandgaps compared to those of their single-element counterparts ($\beta$-As and $\beta$-Sb), while $\beta$-NP has lower scaled bandgap than $\beta$-P. To elucidate the cause of this bandgap variation between $\beta$-NX and $\beta$-X monolayers, first, we need to emphasize that for single-element monolayers $\beta$-X, the origin of the bandgap is the violation of the reflectional symmetry caused by the buckled structure. However, in $\beta$-NX compound monolayers both the violation of the reflectional symmetry and the mass difference between constitute atoms are responsible for the bandgap. The calculated mass ratios of the constitute atoms ($\frac{m_X}{m_N}$) are 2.2, 5.3, and 8.7 in $\beta$-NP, $\beta$-NAs, and $\beta$-NSb, respectively. Also, based on Table 1, $\beta$-X has higher buckling distance $d$ than $\beta$-NX. Thus, one can conclude that the higher mass difference in $\beta$-NAs and $\beta$-NSb can more than compensate the higher buckling distance in $\beta$-As and $\beta$-Sb, results in a higher bandgap in $\beta$-NAs and $\beta$-NSb in comparison with $\beta$-As and $\beta$-Sb. However, the higher buckling distance in $\beta$-P can overcome the effects of the relatively smaller mass ratio in $\beta$-NP, resulting in higher bandgap for $\beta$-P compared to $\beta$-NP.

The phonon group velocity, as defined by Eq. (1), is a crucial parameter in determining the thermal conductivity of a system. Figures 6 to 8 compare the group velocities of $\beta$-NX compounds with those of their single-element monolayers. Also, Figure 9 shows a comparison between acoustic group velocities of $\beta$-NP, $\beta$-NAs, and $\beta$-NSb. Due to the negligible contribution of the optical phonons toward the total thermal conductivity, we only show acoustic modes in these plots. Figures 6 to 8 show that all acoustic phonon modes of $\beta$-NX compounds alloys exhibit higher group velocities than those of the single-element $\beta$-X monolayers, expect for $\beta$-NP for frequencies higher than 5 THz (Figure 6a). This result is consistent with the above discussion about the stiffing of the acoustic modes in $\beta$-NX compounds relative to $\beta$-X. For instance, in the case of $\beta$-NAs ($\beta$-NSb), the maximum group velocities of the ZA, TA and LA modes are 3.19 (2.52) km/s, 6.03 (4.94) km/s and 11.68 (9.92) km/s, respectively. The corresponding values for the ZA, TA and LA group velocities of $\beta$-As ($\beta$-Sb) are 2.12 (1.39) km/s, 3.12 (2.16) km/s, 4.86 (3.41) km/s, respectively. Comparing these values, we can see that some group velocities in $\beta$-NX compounds can be up to about three times higher than those of $\beta$-X, a contributor factor to the higher thermal conductivity of the $\beta$-NX compound monolayers observed in Figures 2(a)-2(c). In Figures 9(a)-9(c), we compare the acoustic group velocities of $\beta$-NP, $\beta$-NAS and $\beta$-NP, showing that the group velocities generally decrease with an increase in average unit cell atomic mass. Such a variation in group velocities implies that the lattice thermal conductivity of these compounds should change as $k_{\beta-NP} > k_{\beta-NAs} > k_{\beta-NSb}$, which contrasts with the actual trend observed in Figure 2(d), $k_{\beta-NAs} > k_{\beta-NSb} > k_{\beta-NP}$. Thus, the group velocity by itself cannot describe the up-and-down variation of the lattice thermal conductivity in $\beta$-NX compounds with the average atomic mass.

Another important parameter in determining the thermal conductivity of a system which solely depends on the harmonic properties is "scattering phase space ($W$)" which is a measurement of the space available for the three-phonon process allowed by the conservation of the energy [49]. This parameter is related to phonon frequencies as [49]

$$W_\lambda^\pm = \frac{1}{2N} \sum_{\lambda' p''} \begin{Bmatrix} 2(n_0' - n_0'') \\ n_0' + n_0'' + 1 \end{Bmatrix} \frac{\delta(\omega_\lambda \pm \omega_{\lambda'} - \omega_{\lambda''})}{\omega_\lambda \omega_{\lambda'} \omega_{\lambda''}} , \qquad (3)$$

which (+) and (-) correspond to the absorption and emissions processes, respectively. A higher value of $W$ means that more space is available for the three-phonon scattering processes which typically is an indicator of lower phonon lifetimes and lower thermal conductivity [41,50–52]. Figure 10 shows the scattering phase space of the ZA, TA and LA modes in $\beta$-NX compounds. Overall, it can be seen that for all phonon modes the phase space increases from $\beta$-NP to $\beta$-NAs to $\beta$-NSb. Due to the inverse correlation between the phase space and the thermal conductivity discussed before, the variation of $W$ suggests that thermal conductivity changes as $k_{\beta-NP} > k_{\beta-NAs} > k_{\beta-NSb}$ among $\beta$-NX compounds monolayers. However, this trend is not the case based on Figure 2(d) and Figure 3. So, based on our discussion in this subsection, although phonon analysis at the harmonic level can describe the higher thermal conductivity of $\beta$-NX compounds compared to $\beta$-X, it is not sufficient to interpret the trend observed between $\beta$-NP, $\beta$-NAs and $\beta$-NSb.

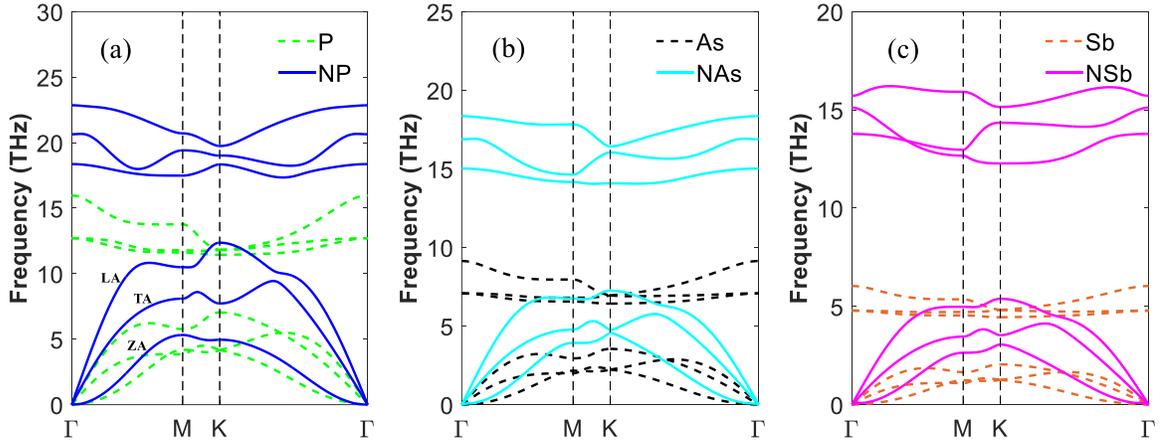

**Fig. 5.** Phonon dispersion curves of (a) $\beta$-NP and $\beta$-P, (b) $\beta$-NAs and $\beta$-As, and (c) $\beta$-NSb and $\beta$-Sb.

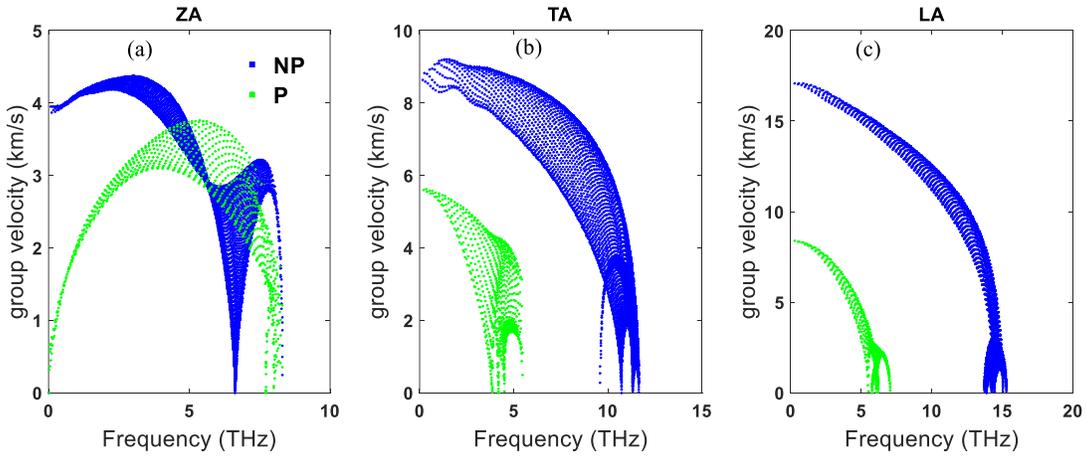

**Fig. 6.** Phonon group velocity of (a) ZA mode, (b) TA mode, and (c) LA mode in β-NP and β-P.

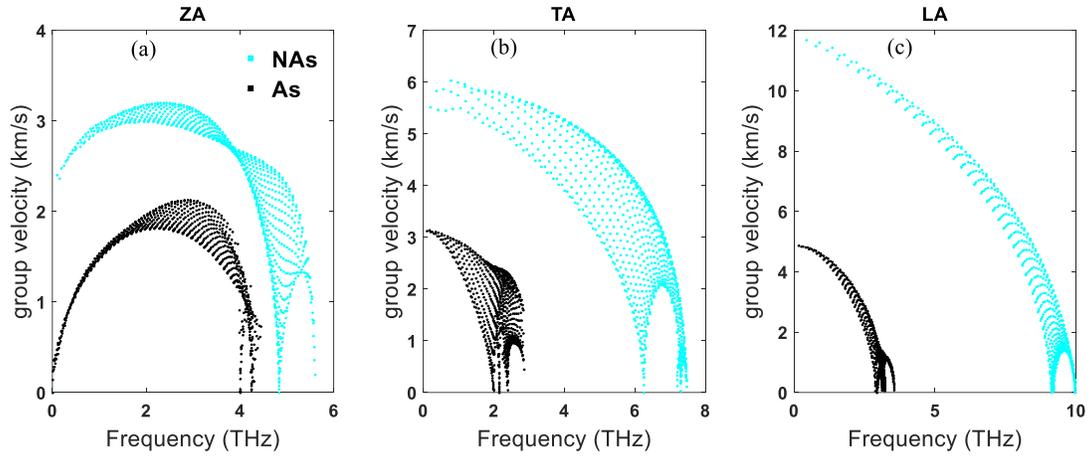

**Fig. 7.** Phonon group velocity of (a) ZA mode, (b) TA mode, and (c) LA mode in β-NAs and β-As.

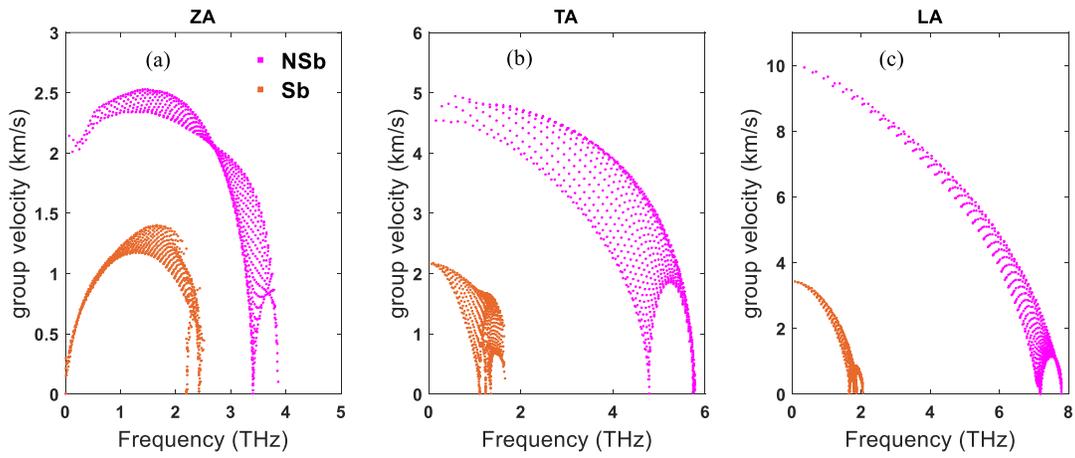

**Fig. 8.** Phonon group velocity of (a) ZA mode, (b) TA mode, and (c) LA mode in β-NSb and β-Sb.

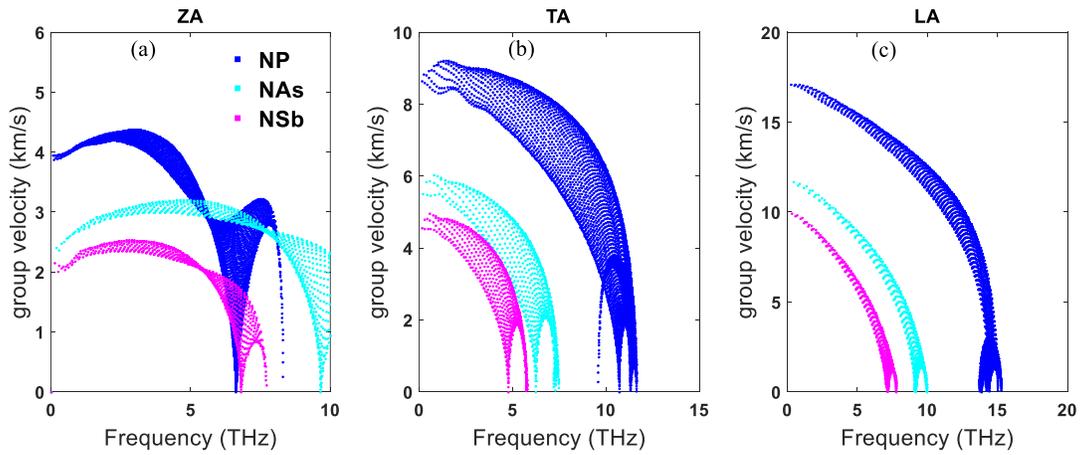

**Fig. 9.** Phonon group velocity of (a) ZA mode, (b) TA mode, and (c) LA mode in β-NP, β-NAs and β-NSb.

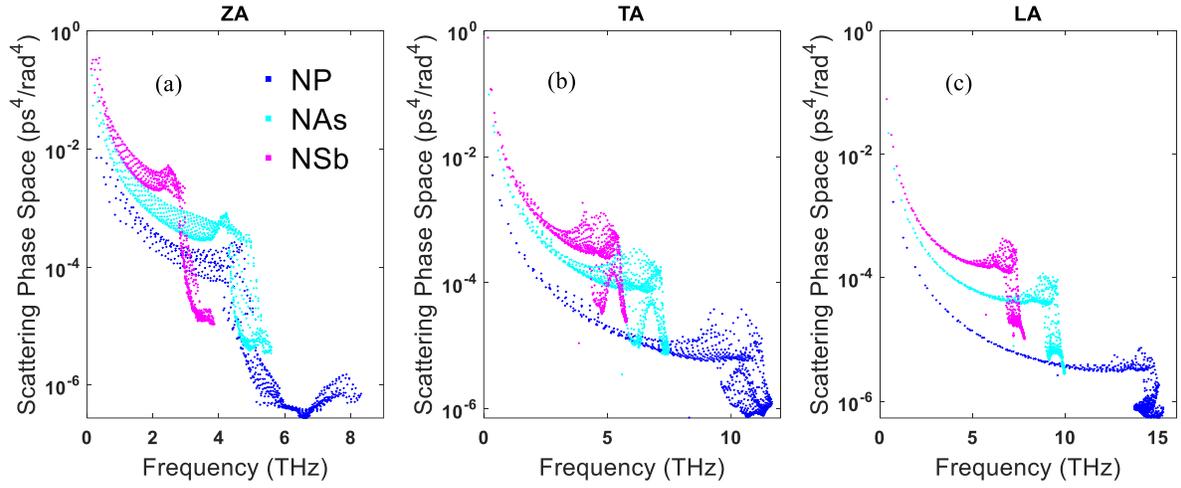

**Fig. 10.** Scattering phase space of the (a) $ZA$, (b) $TA$, and (c) $LA$ modes in $\beta$-NX (X=P, As, Sb) monolayers.

## 3.3- Phonon lifetime and anharmonic properties

Figures 11 to 14 shows the calculated intrinsic three-phonon lifetimes as a function of frequency at 300K for different acoustic phonon modes. Figures 11 to 13 show a comparison between the phonon lifetime of $\beta$-NX compound monolayer and its single element counterpart $\beta$-X. Also, the three-phonon lifetime of $\beta$-NX compounds are compared to each other in Figure 14. Figures 11-13 show that for most of the phonon modes, $\beta$-NX compounds exhibit higher phonon lifetimes than those of their single-element $\beta$-X counterparts. The increase in phonon lifetimes of $\beta$-NX compounds relative to $\beta$-X is more relevant at the low frequency region ($<$ ~1THz), where the difference between phonon lifetimes can be as high as four orders of magnitude. Also, Figures 11-13 show that the higher phonon lifetime of $\beta$-NX compared to $\beta$-X is more relevant in $\beta$-NAs and $\beta$-NSb than $\beta$-NP, which can stem from the higher a-o bandgap in $\beta$-NAs and $\beta$-NSb. The higher phonon lifetime of $\beta$-NX compounds, in almost the whole frequency range, together with its increased group velocities (See Figures 6 to 9), can well justify the higher thermal conductivity of $\beta$-NX compounds compared to those of $\beta$-X. Figures 14(a) to 14(c) compares the acoustic phonon lifetimes of $\beta$-NX compounds. As shown, for most of the phonon modes, phonon lifetimes generally increase from $\beta$-NP to $\beta$-NAs to $\beta$-NSb. Considering the LA mode, we found that at the low frequency region($< \sim$4THz), $\beta$-NAs has the highest phonon lifetime. However, at higher frequencies ($> \sim$4THz), $\beta$-NAs and $\beta$-NSb have approximately similar LA phonon lifetimes, and $\beta$-NP has the lowest. The trends we observed here for phonon lifetimes are similar to those reported for $La_3Cu_3X_4 (X = P, As, Sb, Bi)$ compounds [41], which show that the heaviest system ($La_3Cu_3Bi_4$) has the longest lifetime, whereas the lightest system ($La_3Cu_3P_4$) has the shortest. There are two important factors in determining the phonon lifetime of a system: (i) scattering phase space, and (ii) system anharmonicities. Based on Figure 10 and our discussion in subsection 3.2, $\beta$-NP has the minimum phase space avaible for the three phonon scattering processes and $\beta$-NSb has the highest one. This will give rise to larger phonon lifetimes for $\beta$-NP compared to $\beta$-NAs and $\beta$-NSb. So, exploring the scattering phase space is not enough to understand the underlying reason of the trend we observed for the lifetimes of $\beta$-NX compounds. System anharmonicities, which determines the strength of the three-phonon scattering processes is the key feature that remains to be analyzed. A system with higher anharmonicities has stronger phonon-phonon interactions which give rise to lower phonon lifetime and thermal conductivity. The Grüneisen parameters ($\gamma_\lambda$) [53] is usually used as a measurement of the anharmonicities of a lattice structure. This parameter relates to the anharmonic IFCs by

$$\gamma_\lambda = -\frac{1}{6\omega_\lambda^2} \sum_{ijk\alpha\beta\gamma} \frac{\epsilon_{i\alpha}^{\lambda^*} \epsilon_{j\beta}^{\lambda}}{\sqrt{M_i M_j}} r_k^\gamma \Phi_{ijk}^{\alpha\beta\gamma} e^{i\mathbf{k}.\mathbf{r}_j}, \qquad (4)$$

where α, β and γ are the Cartesian components; $i$, $j$ and $k$ denote atomic indices; $M_i$ represents the mass of atom $i$; $\epsilon_{i\alpha}^{\lambda}$ is the phonon eigenvector for atom $i$ in direction $\alpha$; $\mathbf{r}_i$ is the position vector of $i^{\text{th}}$ atom, and finally $\Phi_{ijk}^{\alpha\beta\gamma}$ represents the third-order anharmonic IFCs. A larger $\gamma_\lambda$ (magnitude) depicts that the system is more anharmonic which results in a lower thermal conductivity. Figure 15(a) shows the mode-dependent Grüneisen parameters for $\beta$-NX compound monolayers. The inset of Figure 15(b) shows the positive values of $\gamma_\lambda$. For most of the phonons, $\beta$-NP and $\beta$-NSb have the highest and lowest Grüneisen parameters, respectively. The trend of $\gamma_\lambda$ observed here is similar to that in $La_3Cu_3X_4(X = P, As, Sb, Bi)$, with $La_3Cu_3P_4$ and $La_3Cu_3Bi_4$ exhibiting the highest and the lowest $\gamma_\lambda$, respectively. Analyzing the Grüneisen parameters combined with the group velocities and the scattering phase space one can well understand the thermal conductivity variation among the $\beta$-NX compounds. There is a competition between harmonic (group velocity and scattering phase space) and anharmonicity of the lattice in determining the thermal conductivity of $\beta$-NX compounds. Harmonic effects tend to decrease the lattice thermal conductivity from $\beta$-NP to $\beta$-NAs to $\beta$-NSb, while higher anharmonicity in $\beta$-NP relative to $\beta$-NAs and $\beta$-NSb compensates by large the relative effects of harmonic properties, giving rise to a lower thermal conductivity for $\beta$-NP. These combined effects result in an up-and-down behavior for the lattice thermal conductivity of $\beta$-NP to $\beta$-NAs to $\beta$-NSb. Interestingly, similar interplays between harmonic and anharmonic properties cause $La_3Cu_3Sb_4$ to have the highest thermal conductivity between $La_3Cu_3X_4(X = P, As, Sb, Bi)$ compounds [41].

## 3.1- Phonon mean free path

Phonon mean free path (MFP) is a critical property when designing nanostructures and controlling the lattice thermal conductivity of a system, which can be measured using the thermal conductivity spectroscopy technique [54]. The normalized thermal conductivity accumulation for $\beta$-NX compounds as well as single element $\beta$-X at 300K are plotted in Figure 16. The normalized thermal conductivity accumulation increases with increasing MPF, and finally converges to one. In order to compare different monolayers, we calculate a critical MPF at which thermal conductivity has 85% of its saturated value. The values of this critical MPF are 35nm, 109 nm, 191 nm, 429 nm, 660 nm, and 690 nm for $\beta$-Sb, $\beta$-As, $\beta$-P, $\beta$-NP, $\beta$-NAs, and $\beta$-NSb, respectively. It can be seen that critical MPF is much higher for all $\beta$-NX compounds compared to their single element counterpart which can stem from higher group velocity and phonon lifetime in $\beta$-NX compounds compared to $\beta$-X. This implies that the effects of sample size on thermal conductivity is more important in $\beta$-NX compounds than single elements group-VA monolayers. Strain engineering is also found to give rise to the size effects in 2D nanostructures [37,55].

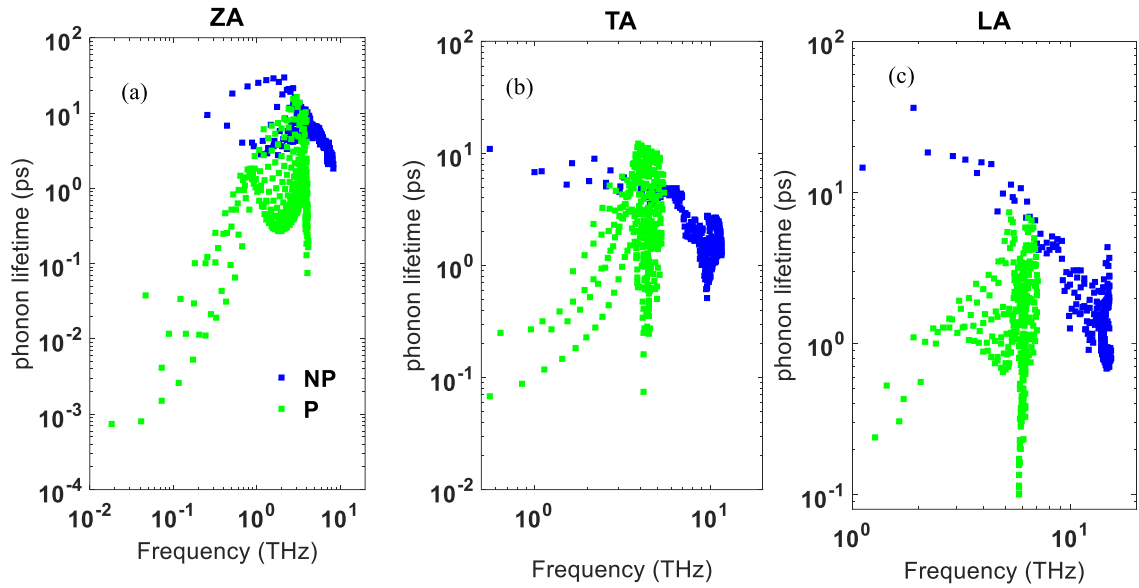

**Fig. 11.** Phonon lifetime of (a) ZA mode, (b) TA mode, and (c) LA mode in β-NP and β-P.

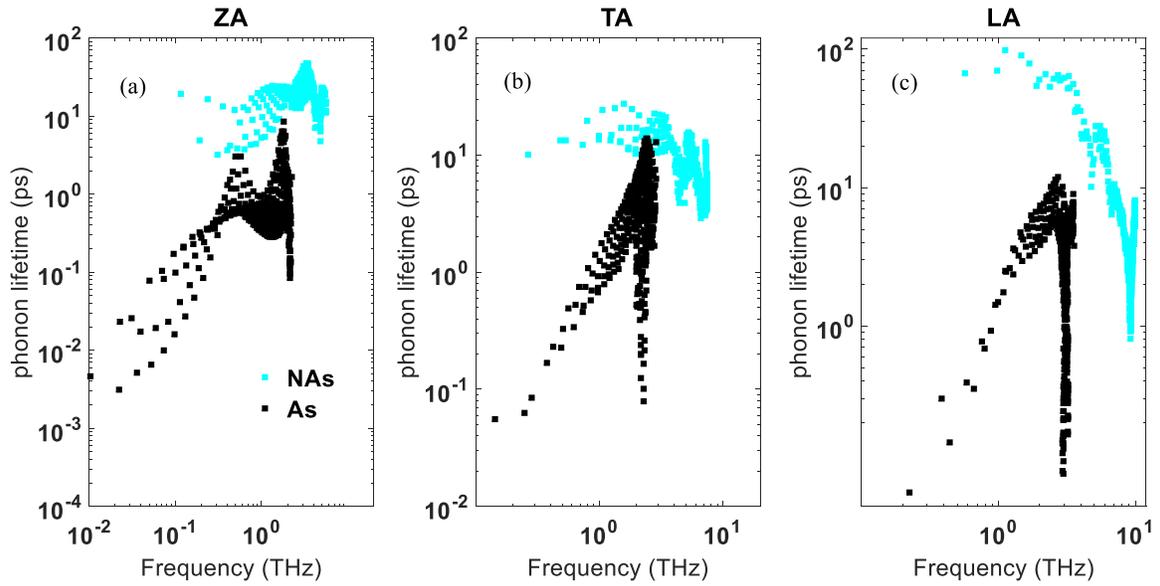

**Fig. 12.** Phonon lifetime of (a) ZA mode, (b) TA mode, and (c) LA mode in β-NAs and β-As.

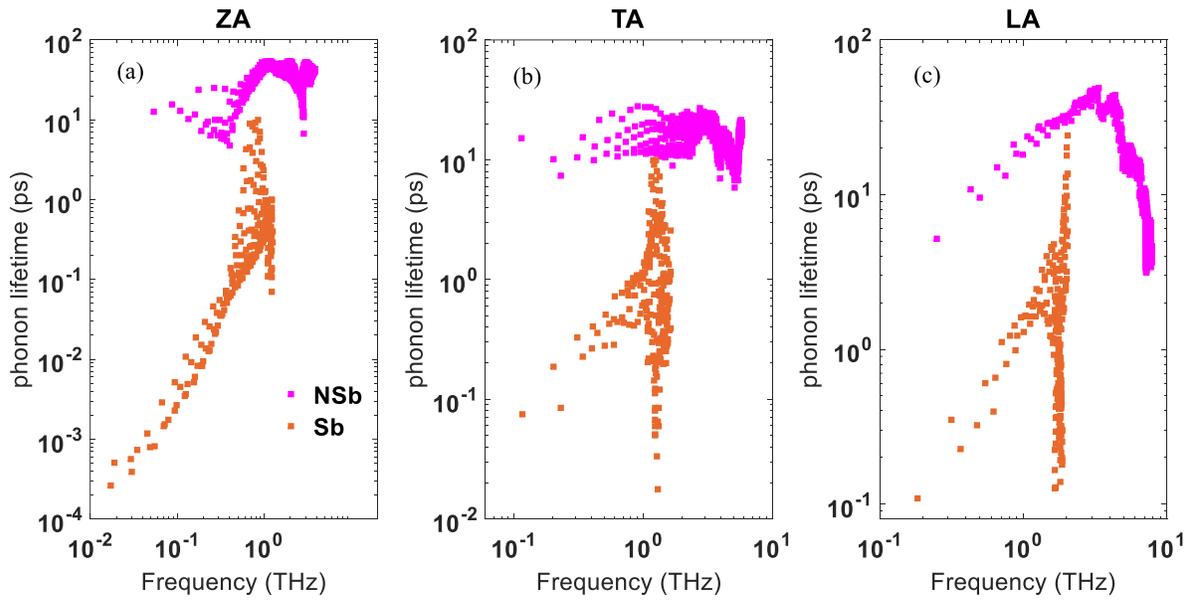

**Fig. 13.** Phonon lifetime of (a) ZA mode, (b) TA mode, and (c) LA mode in β-NSb and β-Sb.

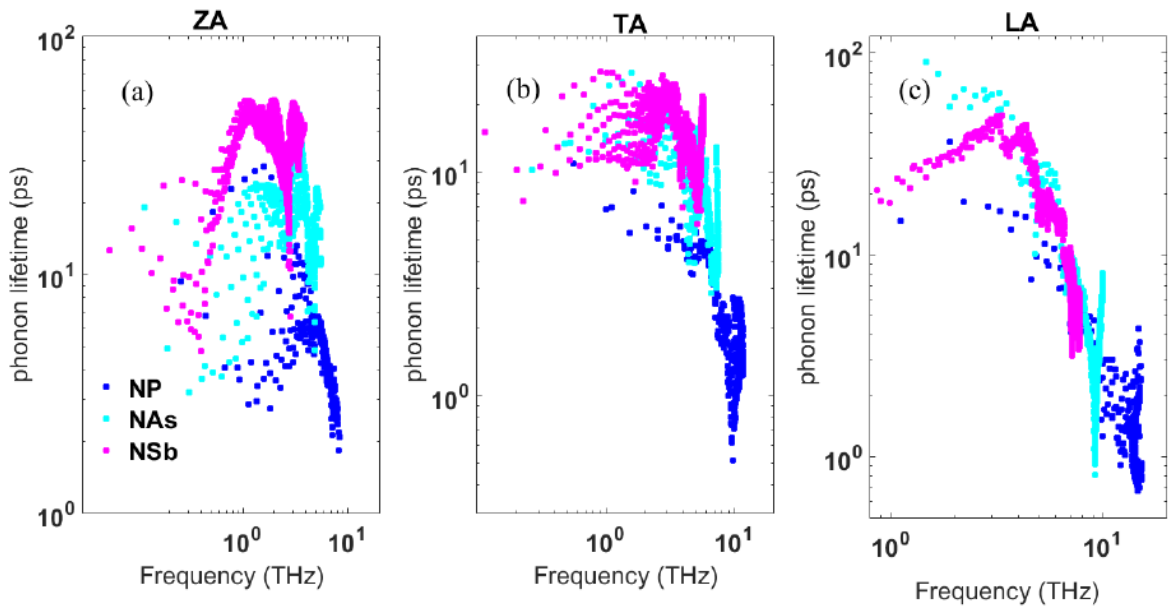

**Fig. 14.** Phonon lifetime of (a) ZA mode, (b) TA mode, and (c) LA mode in β-NP, β-NAs and β-NSb.

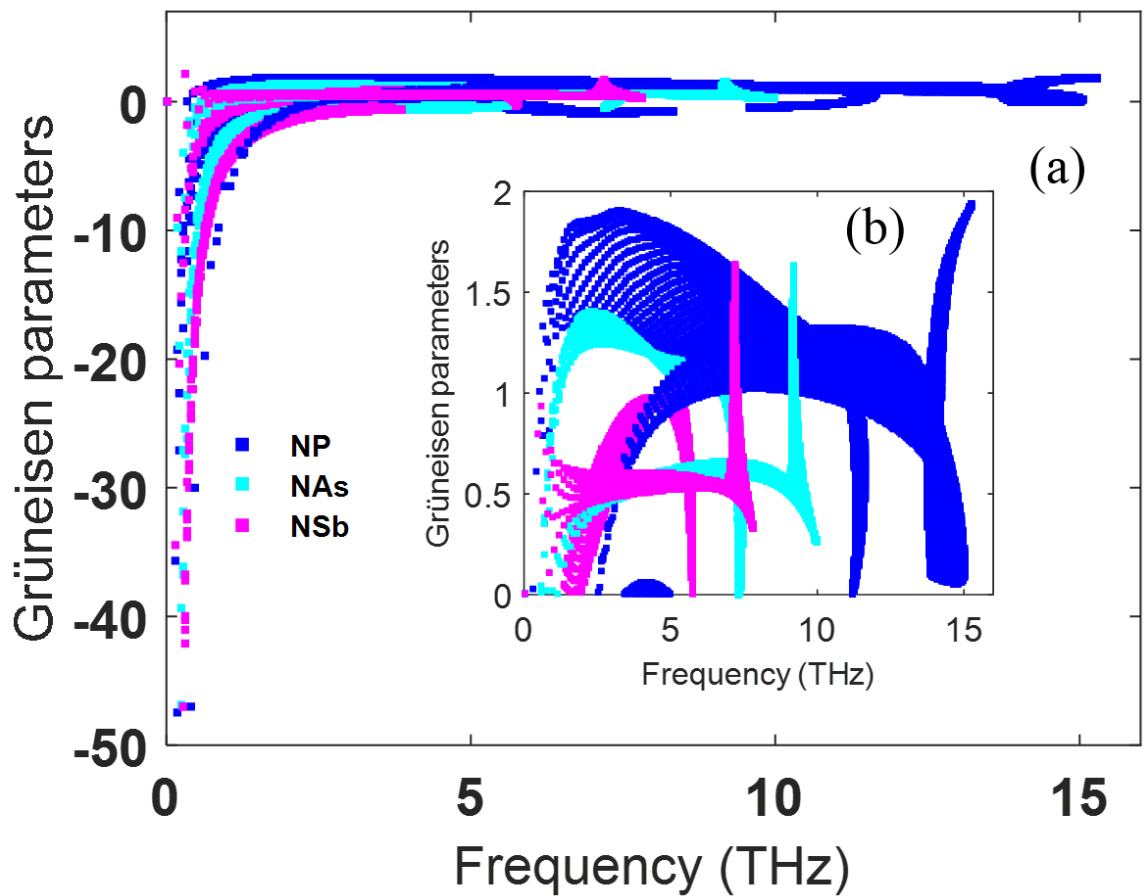

**Fig. 15.** (a) Grüneisen parameters in β-NX (X=P, As, Sb) monolayers (b) Positive values of the Grüneisen parameters.

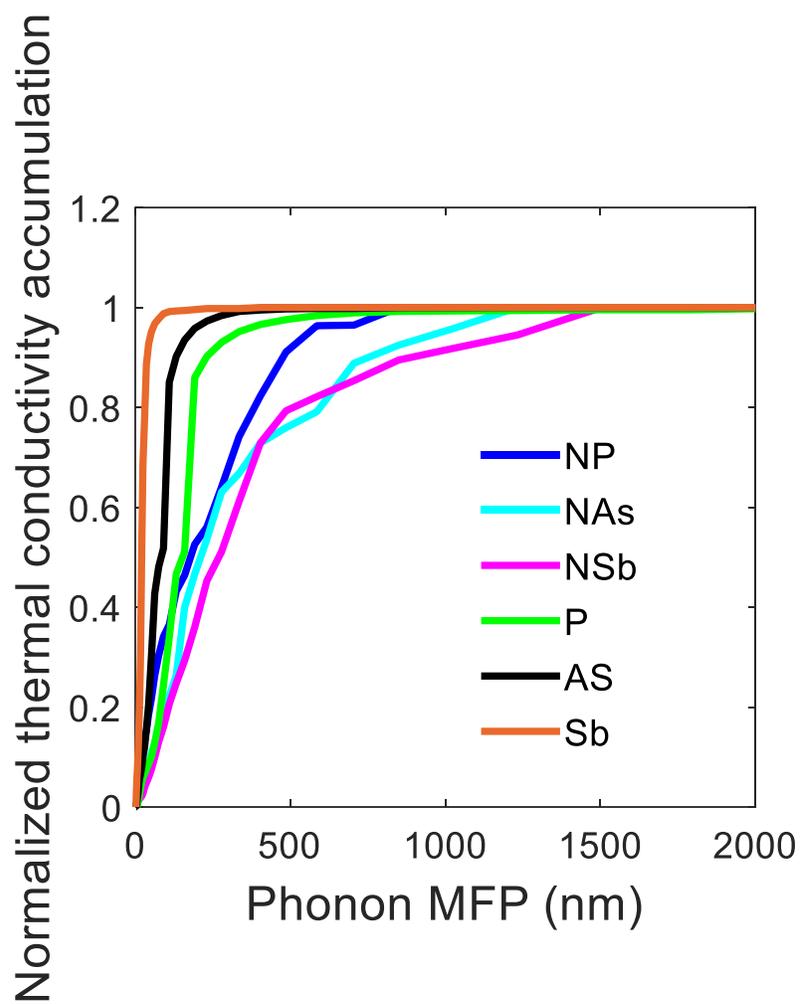

**Fig. 16.** Normalized thermal conductivity accumulation at 300K for β-NX (X=P, As, Sb) and β-X.

# 4-Summary and conclusion

We conducted a rigorous first-principles DFT study based on the iterative solution of the PBTE to investigate phonon thermal transport and the thermal conductivities of novel $\beta$-NX (X=P, As, Sb) compounds monolayers. We found that the room temperature thermal conductivities of 20Å thick layers of $\beta$-NP, $\beta$-NAs and $\beta$-NSb are 28.615 W/mK, 38.46 W/mK, and 34.66 W/mK, respectively. For each of these compound monolayers, the room temperature thermal conductivities are higher than those of their single-element counterparts $\beta$-P, $\beta$-As and $\beta$-Sb. The higher thermal conductivities of $\beta$-NX compared to those of $\beta$-X are attributed to the higher phonon lifetimes and group velocities of $\beta$-NX. Interestingly, we found that despite having the lowest average atomic mass, $\beta$-NP has the minimum thermal conductivity among $\beta$-NX compounds. The analysis of harmonic properties such as group velocity and scattering phase was not sufficient to clarify the thermal conductivity trends observed in $\beta$-NX monolayers. Analyzing the Grüneisen parameter as a measurement of the strength of anharmonic three-phonon processes shows that $\beta$-NP has the highest anharmonicity, which leads to lower phonon lifetime and thermal conductivity. The interplay between harmonic and anharmonic properties is responsible for an up-and-down variation of the lattice thermal conductivity with average atomic mass in $\beta$-NX compound monolayers. This paper highlights the importance of having a comprehensive understanding of the interplay between harmonic and anharmonic phonon properties, providing useful insights into phonon thermal transport of novel 2D nanostructures.

# Conflicts of interest

There is no conflicts to declare.

# Acknowledgment

This work was supported by the Natural Sciences and Engineering Research Council of Canada (NSERC), through the Discovery and the Collaborative Research and Development (CRD) grant programs. Computations were performed on the GPC supercomputer at the SciNet HPC Consortium. SciNet is funded by the Canada Foundation for Innovation under the auspices of Compute Canada; the Government of Ontario; Ontario Research Fund—Research Excellence; and the University of Toronto.